\documentclass[pra,aps,twocolumn,amsmath,amssymb,showpacs]{revtex4}

\begin{document}
\clearpage
\preprint{}

\title{Comment on "Uncertainty relations for positive-operator-valued
measures"}

\author{Alexey E. Rastegin}
\affiliation{Department of Theoretical Physics, Irkutsk State
University, Gagarin Bv. 20, Irkutsk 664003, Russia}

\begin{abstract}
The principal aim of this Comment is to correct those entropic uncertainty
relations that are presented in a paper by Massar [arXiv:quant-ph/0703036v2
(current version)], concerning two approaches to a study of the noise
produced by POVM's. It is next emphasized that the first of the entropic
bounds for POVM obtained by the above author has been already presented in
Ref. [8]. Some exposition obscurity with equation (14) of the commented paper
is elucidated. Finally, some more specific remarks on the paper content are given.
\end{abstract}


\maketitle

\pagenumbering{arabic}
\setcounter{page}{1}

Considering the powers and limitations of generalized measurements,
the writer of Ref. \cite{massar} used the two approaches to the
question. In the second approach, some bounds on the Shannon entropy
of POVM outcomes were examined. Massar's idea to strengthen entropic
bounds by means of the Naimark extension appears to be key original
contribution of Ref. \cite{massar} on the subject of entropic
uncertainty relations. But the two relations presented as Eqs. (16)
and (17) of Ref. \cite{massar} demand corrections. In the
following, the notation of Ref. \cite{massar} will be used.
The Naimark theorem asserts that one can extend the system Hilbert space
in such a way that a generalized measurement has been realized in the
extended space as projective measurement (for details, see \cite{peres}
and references therein). In Ref. \cite{massar} the following
construction is utilized. Let ${\cal{M}}=\{|m_k\rangle\langle m_k|\}$
and ${\cal{N}}=\{|n_l\rangle\langle n_l|\}$ be two POVM's, whose
elements are all rank 1. An extended space
$\widetilde{\cal{H}}=\cal{H}\oplus\cal{H}'$ is the direct sum of the
system space $\cal{H}$ (on which the elements of $\cal{M}$ act) and
an ancillary space $\cal{H}'$. There exists an orthonormal basis of
the extended space $\{|\tilde{m}_k\rangle\}$, which restricted to the
system space gives the POVM $\cal{M}$:
$$
|\tilde{m}_k\rangle=|{m}_k\rangle+|{m}_k'\rangle \ ,
$$
where $|{m}_k\rangle\in\cal{H}$ and $|{m}_k'\rangle\in\cal{H}'$. Massar
points out that the well-known relation (see Eq. (3) of Ref. \cite{massar}),
conjectured by Kraus \cite{kraus} and then established
by Maassen and Uffink \cite{maass}, is directly generalized to such POVM's
\cite{note1}. Here Massar refers to the paper by Hall \cite{hall2}. But Hall
also poses this correct statement without discussion. However, an application
of the Riesz theorem is connected with one delicate aspect. Namely, in the theorem
precondition for transformation the inequality between norms should be valid for
each vector of input Hilbert space \cite{rast2}. Nevertheless, this is corret.
Massar has further observed that the relation can be strengthened by maximizing
a bound over all the possible extensions. His first strengthened relation (see Eq.
(16) of Ref. \cite{massar}) is
$$
H({\cal{M}})+H({\cal{N}})\geq \underset{U'}{\max}-\log_2
\underset{kl}{\max}|\langle\tilde{m}_k|U'|\tilde{n}_l\rangle| \ .
$$
The last inequality should be replaced by the corrected relation
\begin{equation}
H({\cal{M}})+H({\cal{N}})\geq 2\>\underset{U'}{\max}-\log_2
\underset{kl}{\max}|\langle\tilde{m}_k|U'|\tilde{n}_l\rangle|
\ . \label{corr16}
\end{equation}
Here $H({\cal{\cdot}})$ denotes the Shannon entropy of probability
distribution generated by measurement. Putting two different Naimark
extensions of one and the same POVM, Massar then writes down (see
Eq. (17) of Ref. \cite{massar})
$$
H({\cal{M}})\geq \underset{U'}{\max}-\frac{1}{2}\>\log_2
\underset{kl}{\max}|\langle\tilde{m}_k|U'|\tilde{m}_l\rangle| \ .
$$
It is now clear that the last inequality should be replaced by the
corrected relation
\begin{equation}
H({\cal{M}})\geq \underset{U'}{\max}-\log_2
\underset{kl}{\max}|\langle\tilde{m}_k|U'|\tilde{m}_l\rangle|
\ . \label{corr17}
\end{equation}
The two entropic relations given by Eqs. (1) and (2) of this Comment
provide the true results of sharpening the Maassen-Uffink bound by
maximization over Naimark's extensions. These corrected relations
must be used instead of Eqs. (16) and (17) of Ref. \cite{massar}
respectively.

In Section IV of the paper, which is devoted to entropic uncertainty
relations, the first bound for a single POVM (Eq. (13) of Ref.
\cite{massar}) has been previously given by Krishna and Parthasarathy
(see their remark to Corollary 2.6 of Ref. \cite{krishna}). The
written forms of inequality are somewhat distinct, but they are
establishing one and the same entropic bound \cite{note2}.
In addition, some obscurity has been observed in Section IV of Ref.
\cite{massar}. In the paragraph, following Eq. (13) of Ref. \cite{massar},
the author consider a POVM realized by carrying out one of two non-degenerate
projective-valued measures (PVM's)
${\cal{A}}=\{|a_k\rangle\langle a_k|\}$ and
${\cal{B}}=\{|b_l\rangle\langle b_l|\}$ with equal probabilities. He states
that the Maassen-Uffink relation then leads to the bound
\begin{equation}
H({\cal{M}})\geq 1-\frac{1}{2}\>\log_2
\underset{kl}{\max}|\langle{a}_k|{b}_l\rangle|^2 \ .
\label{eq3}
\end{equation}
Applying this inequality to POVM from Eq. (9) of Ref. \cite{massar},
Massar has written Eq. (14). Here it is necessary to state explicitly
the following. If the considered POVM has the form
$$
{\cal{M}}=\left\{\frac{1}{2}\>|a_k\rangle\langle a_k|\right\}
\bigcup\left\{\frac{1}{2}\>|b_l\rangle\langle b_l|\right\}
$$
then the corresponding entropies satisfy
$$
H({\cal{M}})=1+\frac{1}{2}\bigl[H({\cal{A}})+H({\cal{B}})\bigr] \ .
$$
Hence we get Eq. (3) of this Comment. The POVM defined in Eq. (9)
of Ref. \cite{massar} has the above form, and the Maassen-Uffink relation
leads to Eq. (14) of Ref. \cite{massar}.

When in Section I of Ref. \cite{massar} different formulations of
the uncertainty principle are compared, Massar points out that the
Maassen--Uffink bound is independent on the quantum state. At the
same time, in the well-known Robertson relation \cite{robert} a
bound on the product of observable variances can vanish even if
these variances are both positive. Here several remarks seem to be
appropriate. First, state-dependent entropic bounds can quite be
obtained (see Refs. \cite{ruiz2,ghirardi} for two-dimensional case
and Refs. \cite{krishna,rast1} for general case). Rather, for
entropic relations the dependence of a bound on the quantum state
is not critical. Namely, if the state-independent entropic bound
is nontrivial then corresponding state-dependent bound is always
nontrivial \cite{rast2}. At the same time, in tasks of quantum
information processing we usually have only partial or no
knowledge about system state. Although in many cases the
state-dependent entropic relation provides more stronger bound,
state-independent forms are more widely applicable. Second, if a
commutator of two operators is (up to a factor) the identity then
these operators are unbounded \cite{wintner}. So, in
finite-dimensional Hilbert space the Robertson bound is inevitably
dependent on the quantum state. Third, an explicit example of a
shortcoming of the Robertson relation would be suitable. In
the clear example mentioned by Larsen \cite{larsen} the variances
of two observables are nonzero but the Robertson bound is zero
too. Referencing presented by Massar seems to be incomplete. So it
is difficult to place his work in the context of current state of
research of the entropic uncertainty relations.


\begin{thebibliography}{99}
\bibitem{note1}
For two non-degenerate observables, Maassen and Uffink built a
norm-preserving transformation and used the Riesz theorem. This scheme
holds for two POVM's with one-rank elements. Changes are needed for
degenerate observables and arbitrary POVM's, when a transformation
does not preserve a norm.
\bibitem{note2}
Using elementary properties of the operator norm, it is easy to check
the identity of the both bounds. We refrein from presenting the
calculations here.
\end{thebibliography}
\end{document}